\documentclass[aps,prl,twocolumn,superscriptaddress,showpacs,preprintnumbers,amsmath,amssymb]{revtex4}
\usepackage{graphicx} 
\usepackage{dcolumn}  
\usepackage{color}

\RequirePackage{xspace}

\graphicspath{{ps}}

\def\gev    {\ensuremath{\,{\rm GeV}}}
\def\gevc   {\ensuremath{\,{\rm GeV}/\mathit{c}}}
\def\gevcd  {\ensuremath{\,{\rm GeV}/\mathit{c}^2}}
\def\gevdcd {\ensuremath{\,{\rm GeV}^2/\mathit{c}^2}}
\def\gevdcq {\ensuremath{\,{\rm GeV}^2/\mathit{c}^4}}

\def\btou{\ensuremath{b \to u}\xspace}
\def\btoc{\ensuremath{b \to c}\xspace}

\def\Bxulnu{\ensuremath{B \to X_u \ell \nu}\xspace}
\def\Bxclnu{\ensuremath{B \to X_c \ell \nu}\xspace}

\def\Vubs{\ensuremath{|V_{ub}|}\xspace}

\def\ysters{\ensuremath{\Upsilon(4S)}}
\def\Bbar{\kern 0.18em\overline{\kern -0.18em B}{}\xspace}

\def\eeqq{\ensuremath{e^+e^- \to q\bar{q}}}

\def\mbSF    {\ensuremath{m_b(\rm SF)}}
\def\mupidSF {\ensuremath{\mu_{\pi}^2(\rm SF)}}
\def\mbSFv   {\ensuremath{\mbSF=(4.60 \pm 0.04)~\gevcd}}
\def\mupidSFv{\ensuremath{\mupidSF=(0.20 \pm 0.04)~\gevdcd}}
\def\LSF     {\ensuremath{\Lambda^{\rm SF}}}
\def\lSF     {\ensuremath{\lambda_1^{\rm SF}}}

\def\onlumi{\ensuremath{253\;{\rm fb}^{-1}}}
\def\oflumi{\ensuremath{28\;{\rm fb}^{-1}}}
\def\noofB {\ensuremath{275 \times 10^6\; B \overline{B}}}

\def\mm    {\ensuremath{m_{\rm miss}}}
\def\mmds  {\ensuremath{m_{\rm miss\;(\mathit{D^*})}}}
\def\epsbu {\ensuremath{\varepsilon_{\rm sel}^{b \to u}}}
\def\Eb    {\ensuremath{E_{\rm beam}}}
\def\polarlcut{\ensuremath{26^{\circ} \leq \theta \leq 140^{\circ}}}
\def\Nsl {\ensuremath{N_{\rm sl}=(9.14 \pm 0.05) \times 10^4}}
\def\fake{\ensuremath{14.0\%}}
\def\Brsl{\ensuremath{{\cal{B}}(B \to X \ell \nu) = 0.1073 \pm 0.0028}}

\def\mmissdstar{\ensuremath{\mmds^2=(p_B-p_{D^*}-p_{\ell})^2}}
\def\mmissdstarcut{\ensuremath{\mmds^2>-3\gevdcq}}

\def\mmisscut {\ensuremath{-1 \leq \mm^2 \leq 0.5\gevdcq}}
\def\thrustcut{\ensuremath{|\cos \theta_{\rm thrust}^*|<0.8}}
\def\Mbccut   {\ensuremath{M_{\rm bc} \geq 5.27 \gevcd}}
\def\DEcut    {\ensuremath{-0.2 < \Delta E < 0.05 \gev}}
\def\sr   {\ensuremath{\Delta \Phi}}
\def\srMx {\ensuremath{M_X < 1.7 ~\gevcd}}
\def\srqd {\ensuremath{q^2 > 8   ~\gevdcd}}
\def\srpp {\ensuremath{P_+ < 0.66~\gevc}}
\def\srps {\ensuremath{p_{\ell}^\ast \geq 1\gevc}}

\def\prate  {\ensuremath{\Delta\Gamma_{u \ell \nu}(\sr)}}

\def\rfrec{\ensuremath{\varepsilon_{\rm frec}^{\rm sl}/\varepsilon_{\rm frec}^{b \to u}}}
\def\rlept{\ensuremath{\varepsilon_{\ell}^{\rm sl}/\varepsilon_{\ell}^{b \to u}}}
\def\r    {\ensuremath{r_{b\to u}^{\rm sl}}}

\def\epsbu {\ensuremath{\varepsilon_{\rm sel}^{b \to u}}}
\def\Nburaw{\ensuremath{N_{b \to u}^{\rm raw}}}

\def\Btag  {\ensuremath{B_{\rm tag}}}
\def\Bsig  {\ensuremath{B_{\rm sig}}}

\def\Vubv{\ensuremath{(4.09 \pm 0.19 \pm 0.20 {}_{\,-0.15}^{\,+0.14} \pm 0.18)\times 10^{-3}}}

\def\Rmxqd{\ensuremath{23.7\pm 2.0({\rm SF}){}_{\,-2.3}^{\,+2.5}({\rm th.})}}
\def\Rmx  {\ensuremath{46.1\pm 4.2({\rm SF}){}_{\,-3.2}^{\,+3.5}({\rm th.})}}
\def\Rppl {\ensuremath{39.4\pm 4.5({\rm SF}){}_{\,-2.7}^{\,+2.8}({\rm th.})}}

\def\tauB{\ensuremath{\tau_B=(1.604\pm 0.016)\;{\rm ps}}}
\def\vubformula{\ensuremath{|V_{ub}|^2 = \prate/R(\sr)}}

\begin{document}

\title{
Determination of \Vubs from Measurements of the Inclusive Charmless
Semileptonic Partial Rates of $B$ Mesons using Full Reconstruction
Tags}

\affiliation{Budker Institute of Nuclear Physics, Novosibirsk}
\affiliation{Chiba University, Chiba}
\affiliation{Chonnam National University, Kwangju}
\affiliation{University of Cincinnati, Cincinnati, Ohio 45221}
\affiliation{Gyeongsang National University, Chinju}
\affiliation{University of Hawaii, Honolulu, Hawaii 96822}
\affiliation{High Energy Accelerator Research Organization (KEK), Tsukuba}
\affiliation{Hiroshima Institute of Technology, Hiroshima}
\affiliation{Institute of High Energy Physics, Chinese Academy of Sciences, Beijing}
\affiliation{Institute of High Energy Physics, Vienna}
\affiliation{Institute for Theoretical and Experimental Physics, Moscow}
\affiliation{J. Stefan Institute, Ljubljana}
\affiliation{Kanagawa University, Yokohama}
\affiliation{Korea University, Seoul}
\affiliation{Kyungpook National University, Taegu}
\affiliation{Swiss Federal Institute of Technology of Lausanne, EPFL, Lausanne}
\affiliation{University of Ljubljana, Ljubljana}
\affiliation{University of Maribor, Maribor}
\affiliation{University of Melbourne, Victoria}
\affiliation{Nagoya University, Nagoya}
\affiliation{Nara Women's University, Nara}
\affiliation{National Central University, Chung-li}
\affiliation{National United University, Miao Li}
\affiliation{Department of Physics, National Taiwan University, Taipei}
\affiliation{H. Niewodniczanski Institute of Nuclear Physics, Krakow}
\affiliation{Nippon Dental University, Niigata}
\affiliation{Niigata University, Niigata}
\affiliation{Nova Gorica Polytechnic, Nova Gorica}
\affiliation{Osaka City University, Osaka}
\affiliation{Osaka University, Osaka}
\affiliation{Panjab University, Chandigarh}
\affiliation{Peking University, Beijing}
\affiliation{Princeton University, Princeton, New Jersey 08544}
\affiliation{Saga University, Saga}
\affiliation{University of Science and Technology of China, Hefei}
\affiliation{Seoul National University, Seoul}
\affiliation{Sungkyunkwan University, Suwon}
\affiliation{University of Sydney, Sydney NSW}
\affiliation{Tata Institute of Fundamental Research, Bombay}
\affiliation{Toho University, Funabashi}
\affiliation{Tohoku Gakuin University, Tagajo}
\affiliation{Tohoku University, Sendai}
\affiliation{Department of Physics, University of Tokyo, Tokyo}
\affiliation{Tokyo Institute of Technology, Tokyo}
\affiliation{Tokyo Metropolitan University, Tokyo}
\affiliation{Tokyo University of Agriculture and Technology, Tokyo}
\affiliation{University of Tsukuba, Tsukuba}
\affiliation{Virginia Polytechnic Institute and State University, Blacksburg, Virginia 24061}
\affiliation{Yonsei University, Seoul}
  \author{I.~Bizjak}\affiliation{J. Stefan Institute, Ljubljana} 
  \author{K.~Abe}\affiliation{High Energy Accelerator Research Organization (KEK), Tsukuba} 
  \author{K.~Abe}\affiliation{Tohoku Gakuin University, Tagajo} 
  \author{H.~Aihara}\affiliation{Department of Physics, University of Tokyo, Tokyo} 
  \author{Y.~Asano}\affiliation{University of Tsukuba, Tsukuba} 
  \author{S.~Bahinipati}\affiliation{University of Cincinnati, Cincinnati, Ohio 45221} 
  \author{A.~M.~Bakich}\affiliation{University of Sydney, Sydney NSW} 
  \author{Y.~Ban}\affiliation{Peking University, Beijing} 
  \author{E.~Barberio}\affiliation{University of Melbourne, Victoria} 
  \author{M.~Barbero}\affiliation{University of Hawaii, Honolulu, Hawaii 96822} 
% ...

\collaboration{The Belle Collaboration}

\begin{abstract}
We present a measurement of the Cabibbo-Kobayashi-Maskawa matrix
element \Vubs, based on \onlumi\ of data collected by the Belle
detector at the KEKB $e^+e^-$ asymmetric collider. Events are tagged
by fully reconstructing one of the $B$ mesons, produced in pairs from
\ysters. The signal for \btou semileptonic decay is distinguished from
the \btoc background using the hadronic
mass $M_X$, the leptonic invariant mass squared $q^2$ and the variable
$P_+ \equiv E_X-|\vec{p}_X|$. The results are obtained
for events with the prompt-lepton momentum, {\srps}, in three
kinematic regions (1)~{\srMx}, (2)~{\srMx} combined with {\srqd}, and by
(3)~{\srpp}, allowing for a comparison of the three methods. The
matrix element \Vubs is found to be \Vubv, where the errors are
statistical, systematic including Monte Carlo modeling, theoretical and from
shape function parameter determination, respectively.
\end{abstract}


\maketitle
\tighten

{\renewcommand{\thefootnote}{\fnsymbol{footnote}}}
\setcounter{footnote}{0}

%
An accurate knowledge of the Cabibbo-Kobayashi-Maskawa matrix element
\Vubs is crucial to test Standard Model predictions for $CP$
violation.
Currently, the best precision may be achieved by measuring the
inclusive rate \prate\ of \Bxulnu\ decays in a restricted
region of the phase space (\sr) where the dominant charm background is
suppressed and theoretical uncertainties are minimized.  The
theoretical factor $R(\sr)$ directly relates the inclusive rate to \Vubs,
with no extrapolation to the full phase space:
$\Vubs^2 = \prate / R(\sr)$. Here we report measurements of \prate\
for several choices of \sr\ and derive the corresponding values of \Vubs.

The measurements are made with a sample of events where the hadronic
decay mode of the tagging side $B$ meson,
\Btag, is fully reconstructed, while the semileptonic decay of the
signal side $B$ meson, \Bsig, is identified by the presence of a high
momentum electron or muon. $B$ denotes both charged and neutral $B$
mesons. This method allows the construction of the invariant masses
of the hadronic ($M_X$) and leptonic ($\sqrt{q^2}$) system in the
semileptonic decay, and the variable $P_{+} \equiv E_X-|\vec{p}_X|$,
where $E_X$ is the energy and $|\vec{p}_X|$ the magnitude of the
three-momentum of the hadronic system.  These inclusive kinematic
variables can be used to separate the \Bxulnu decays from the much
more abundant \Bxclnu decays. Three competing kinematic regions (\sr)
were proposed by theoretical studies~\cite{bib:q2cut,bib:bosch}, based
on the three kinematic variables, and are directly compared by this
analysis. The value of \Vubs is extracted using recent theoretical
calculations~\cite{bib:bosch,bib:generator1} that include all the
currently known contributions. $M_X$ and $q^2$ selections were already
used to to extract \Vubs~\cite{bib:Babar_frec,bib:Kakuno}. The present
analysis is the first one to use $P_{+}$ and to directly compare the
three methods.

The data were collected with the Belle
detector~\cite{bib:BELLE} at the asymmetric-energy KEKB storage
ring~\cite{bib:KEKB}. The results presented in this paper are based on
a \onlumi\  sample recorded at the \ysters\ resonance, which
contains \noofB\ pairs.  An additional \oflumi\  sample taken at a
center-of-mass energy 60\,MeV below the \ysters\ resonance is used to
subtract the background from \eeqq\ ($q=u,\ d,\ s,\ c$).

Monte Carlo (MC) simulated events were used to determine efficiencies
as well as signal and background distributions. The detector
simulation was based on GEANT~\cite{bib:GEANT}.  To model {\Bxulnu} we
use the EvtGen generator~\cite{bib:evtgen} with various models, where
$X_u$ is $\pi$ or $\rho$~\cite{bib:bu-model-lcsr}, an excited $X_u$
state~\cite{bib:bu-model-isgw2}, or a non-resonant multiparticle final
state~\cite{bib:bu-model-fn}. The \Bxclnu transitions are simulated
according to the QQ generator~\cite{bib:bc-model}.  For the two
dominant contributions, $D^* \ell \nu$ and $D \ell \nu$, we use an
HQET-based parametrization of form factors~\cite{bib:HQET-ff} and
ISGW2 model \cite{bib:bu-model-isgw2}, respectively. For the $D^{**}$
we use ISGW2 model and for sub-components $D_1$ and $D_2^*$ set
$\frac{{\cal B} \to D_1 \ell \nu + {\cal B} \to D_2^* \ell \nu}{{\cal B} \to D^{**} \ell \nu}=0.35\pm 0.23$.
The motion of the $b$ quark inside the $B$ meson is implemented with
the introduction of a shape function~\cite{bib:bu-model-fn,bib:generator2}
that describes the $b$ quark momentum distribution inside the $B$
meson.

The \Btag\ candidates are reconstructed in the modes $B \to
D^{(\ast)} \pi$/$\rho$/$a_1$/$D^{(\ast)}_s$, 
$\overline{D}{}^0 \to K^+ \pi^-$, $K^+\pi^- \pi^0$, $K^+ \pi^+ \pi^-
\pi^-$, $K_S^0 \pi^0$, $K_S^0 \pi^+\pi^-$, $K_S^0 \pi^+ \pi^- \pi^0$
and $K^+ K^-$, $D^- \to K^+ \pi^- \pi^-$, $K^+
\pi^- \pi^- \pi^0$, $K_S^0 \pi^-$, $K_S^0 \pi^- \pi^0$, $K_S^0 \pi^-
\pi^- \pi^+$ and $K^+ K^- \pi^-$, and $D_s^+ \to K_S^0 K^+$
and $K^+ K^- \pi^+$. 
$\overline{D}{}^{\ast}$ mesons are reconstructed by combining a
$\overline{D}$ candidate and a soft pion or photon. (Inclusion of
charge conjugate decays is implied throughout this paper.)
The selection of \Btag\ candidates is based on the
beam-constrained mass, $M_{\rm bc} = \sqrt{\Eb^{\ast 2}/c^4 -
p^{\ast 2}_{B}/c^2}$, and the energy difference, $\Delta E =
E^{\ast}_{B} - \Eb^{\ast}$. Here $\Eb^{\ast}=\sqrt{s}/2
\simeq 5.290 \gev$ is the beam energy in the $e^+e^-$ center-of-mass
system (cms), and $p^{\ast}_B$ and $E^{\ast}_B$ are the cms momentum
and energy of the reconstructed $B$ meson. (Throughout this paper the
variables calculated in the cms are denoted with an asterisk.)

The combinatorial background from jet-like \eeqq\ processes is
suppressed by an event topology requirement based on the normalized
second Fox-Wolfram moment $R_2<0.5$~\cite{bib:R2}, and for some modes
also by \thrustcut, where $\theta_{\rm thrust}^*$ is
the angle between the thrust axis of the \Btag\ candidate and that of
the rest of the event.
To minimize the fraction of events with incorrect separation of tag
and signal sides while maintaining high signal efficiency, a loose
selection requirement of $M_{\rm bc} \geq 5.22 \gevcd$ and {\DEcut} is
made.
If an event has multiple \Btag\ candidates, we choose the one having
the smallest $\chi^2$ based on $\Delta E$, the $D$ candidate mass, and
the $D^{\ast}-D$ mass difference if applicable.

\begin{figure}[t]
\centerline{
\includegraphics[width=0.25\textwidth]{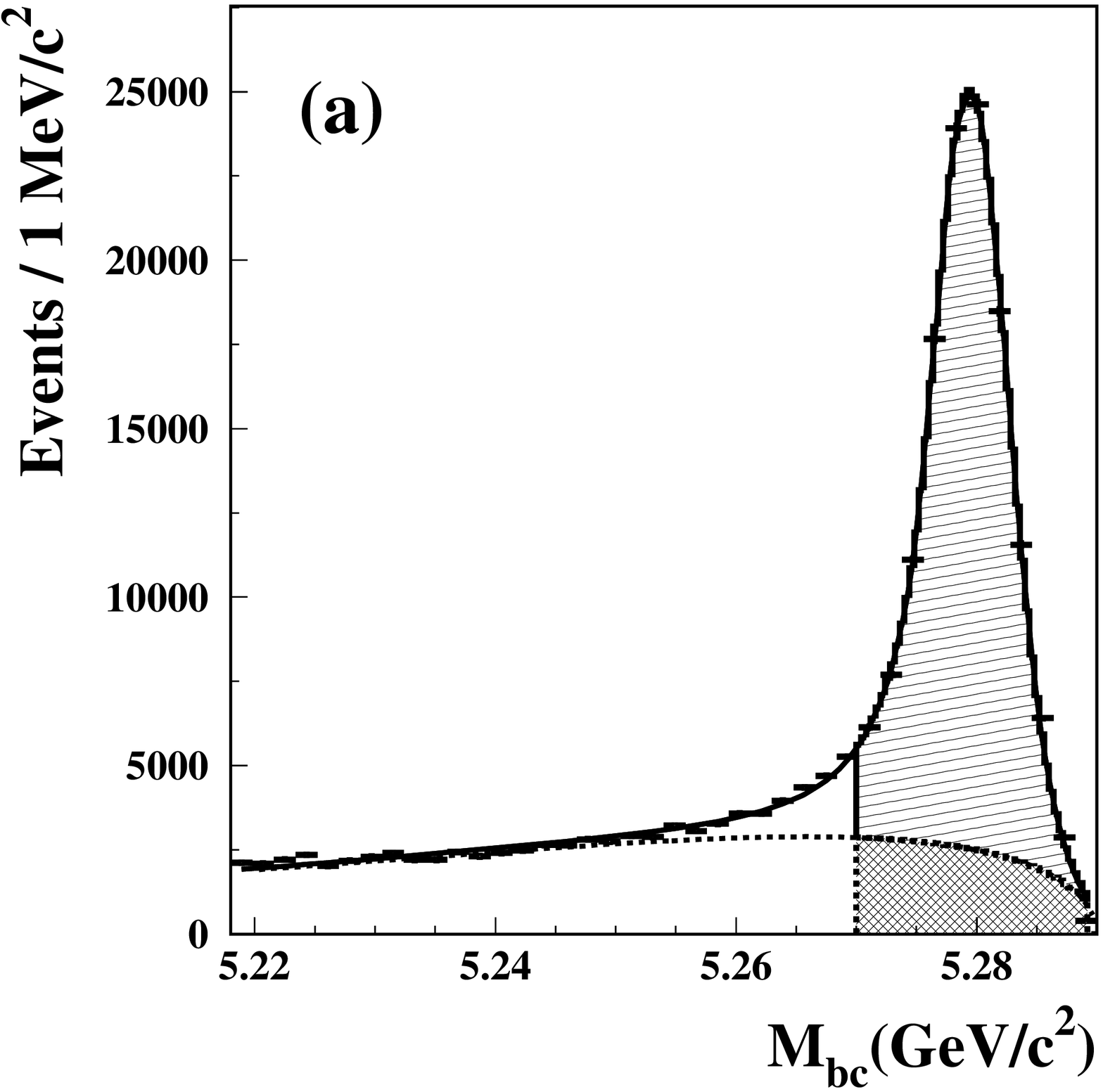}
\includegraphics[width=0.25\textwidth]{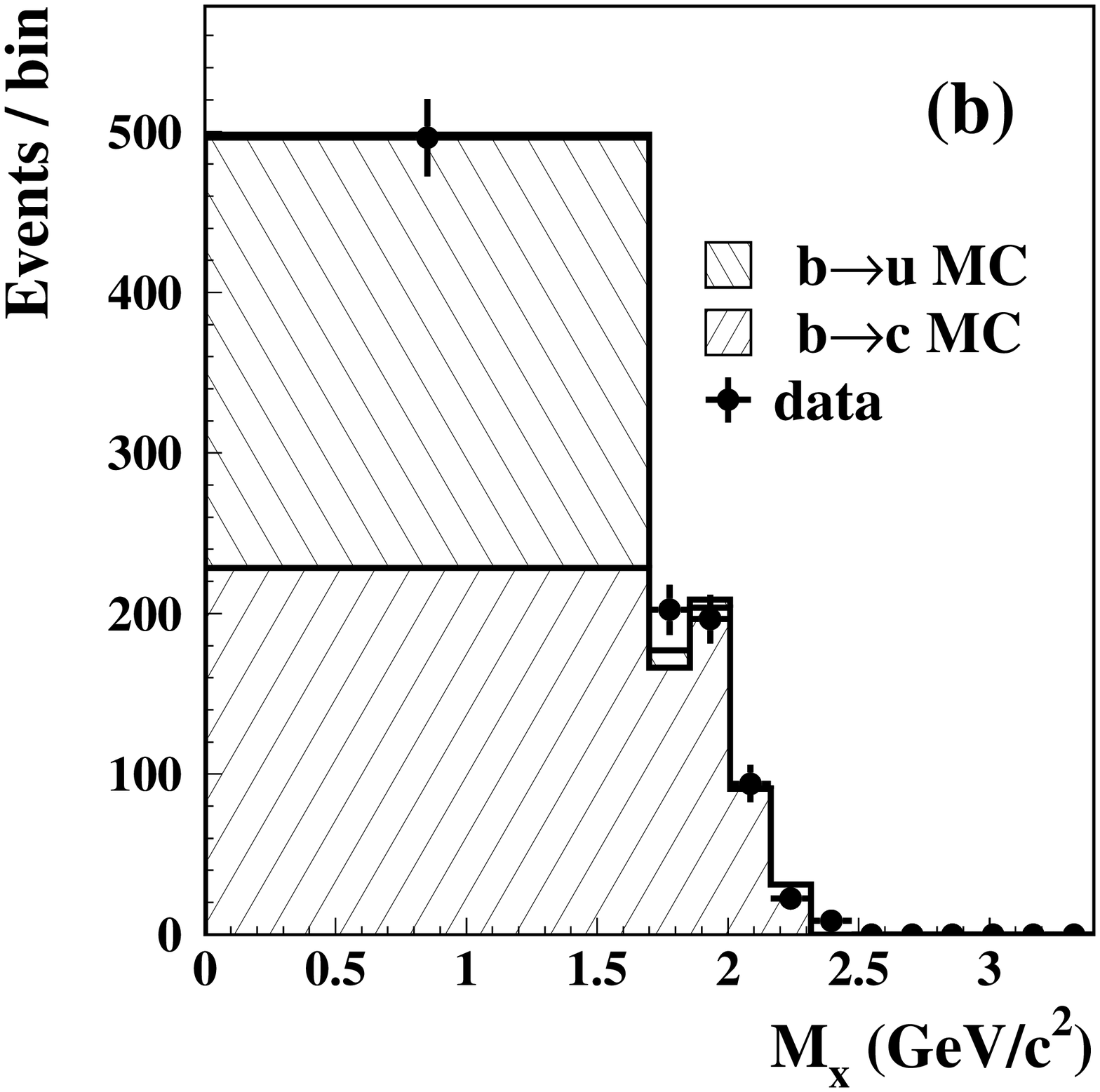}}
\caption{
(a) Distribution in $M_{\rm bc}$ (data) of {\Btag}
candidates in events satisfying {\Bsig}
selection. 
(b) $M_X$ distribution for events with
{\srqd}, with fitted contributions of \Bxclnu and {\Bxulnu}.}
\label{fig:Mbc-Mxq2} 
\end{figure}

For events tagged by fully reconstructed \Btag\ candidates, we search
for electrons or muons from semileptonic decays of {\Bsig}. We require
a lepton with momentum $p_{\ell}^{\ast}$ exceeding $1\gevc$ in the
laboratory polar angular region of \polarlcut.
Leptons from $J/\psi$ decay, photon conversion in the material of the
detector, and $\pi^0$ decay are rejected based on the invariant mass
they form in combination with an oppositely charged lepton and for
electron candidates also with an additional photon.  When the \Btag\
candidate is charged, we also require the lepton charge to be
consistent with that from prompt semileptonic decay. 
The signal yield is obtained by fitting the $M_{\rm bc}$ distribution
to the sum of an empirical parametrization of the combinatorial
background shape~\cite{bib:argus} plus a signal shape~\cite{bib:cball}
that peaks at the $B$ mass and taking the part of the signal that lies
in the ``signal region,'' {\Mbccut}, as shown in
Fig.~\ref{fig:Mbc-Mxq2}(a).  The cutoff for $M_{\rm bc}$ reduces the
uncertainty from the incorrect assignment of tag and signal sides in
signal events.

The \Bxulnu signal events are selected by removing poorly measured
soft charged tracks and imposing several additional
requirements to reject poorly reconstructed events and suppress the
\Bxclnu background.  We require that the event contain exactly one
lepton and have zero net charge and that the invariant mass squared of
the missing four-momentum $\mm^2\equiv (p_{\ysters} - p_{\Btag} -
p_{X} - p_\ell)^2$ ($p_{\ysters}$, $p_{\Btag}$ and $p_X$ are
four-momenta of the {\ysters}, \Btag, and hadronic system ($X$),
respectively) be within \mmisscut. To suppress the \Bxclnu
background, events with a $K^\pm$ or $K_S^0$ candidate on the signal
side are rejected (kaon veto).
To reject events containing a $K_L^0$, we require that
the angle between the missing momentum and the direction of any
$K_L^0$ candidate, reconstructed in the $K_L^0$ detector, be greater
than 37 degrees. We also reject $B^0 \to D^{\ast +} \ell^- \bar\nu$
events by detecting the slow pion ($\pi_s$) from $D^{\ast +} \to D^0
\pi_s^+$ and deducing from its momentum the momentum of the $D^{\ast
+}$. The missing mass squared \mmissdstar\ is calculated from the
reconstructed quantities, and events with \mmissdstarcut\ are
rejected.

\begin{figure}[t]
\centerline{
\includegraphics[width=0.25\textwidth]{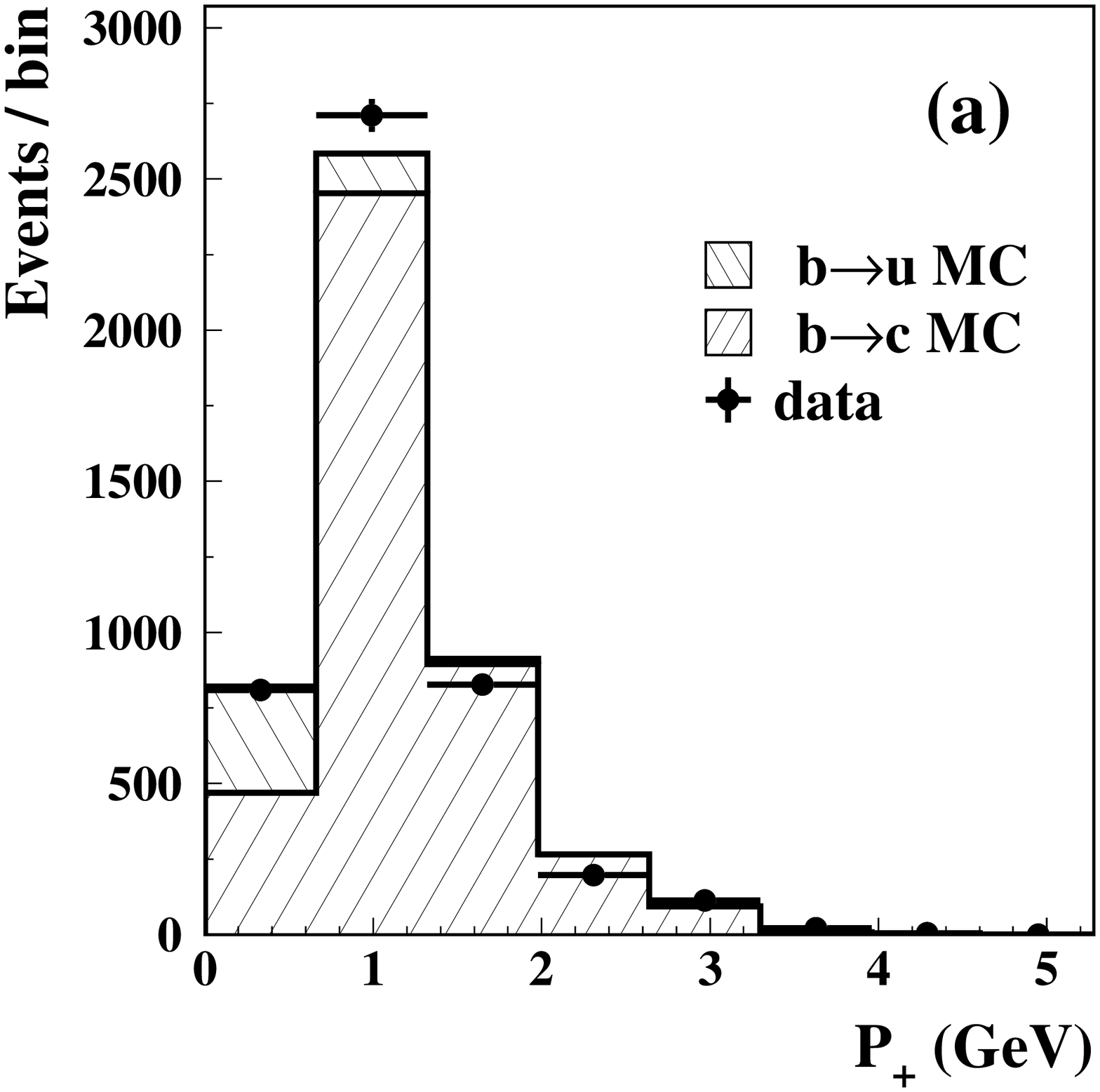}
\includegraphics[width=0.25\textwidth]{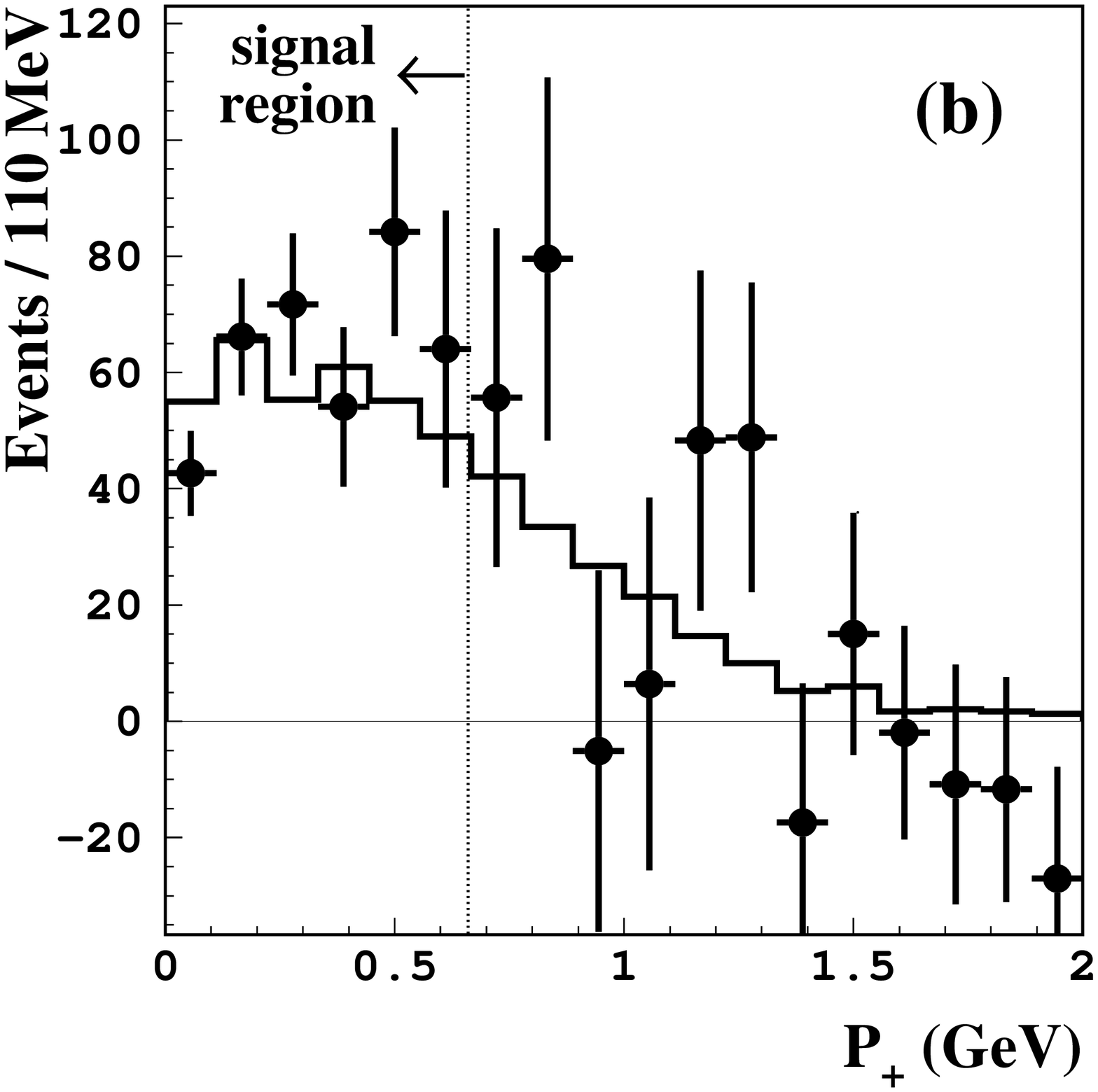}}
\caption{
(a) The $P_+$ distribution for the selected events, with fitted
contributions from \Bxclnu and \Bxulnu, (b) $P_+$ distribution
(symbols with error bars) after subtracting \Bxclnu, with fitted $B
\to X_u \ell \nu$ contribution (histogram).}
\label{fig:ppl}
\end{figure}

\begin{figure}
\centerline{
\includegraphics[width=0.25\textwidth]{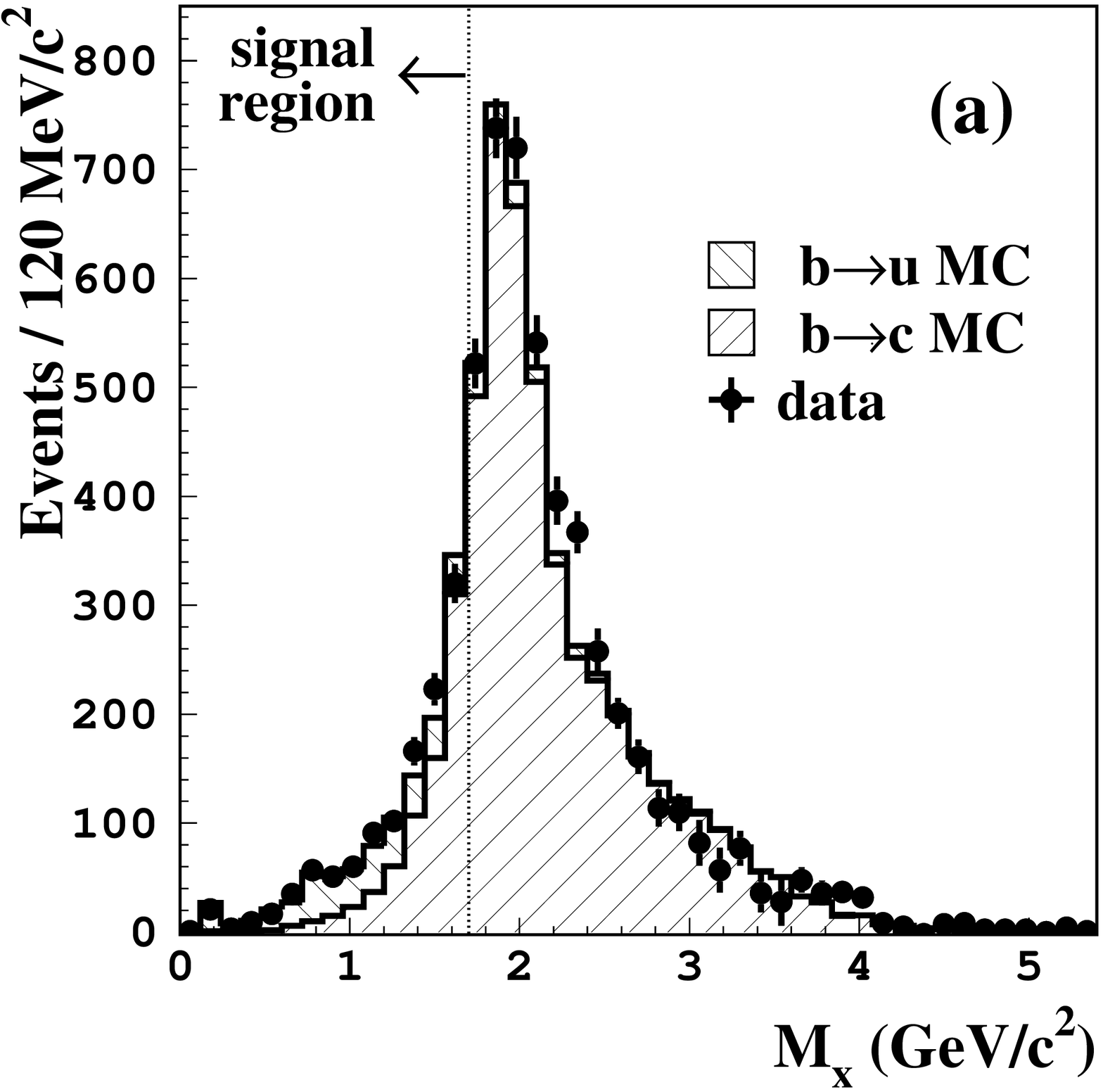}
\includegraphics[width=0.25\textwidth]{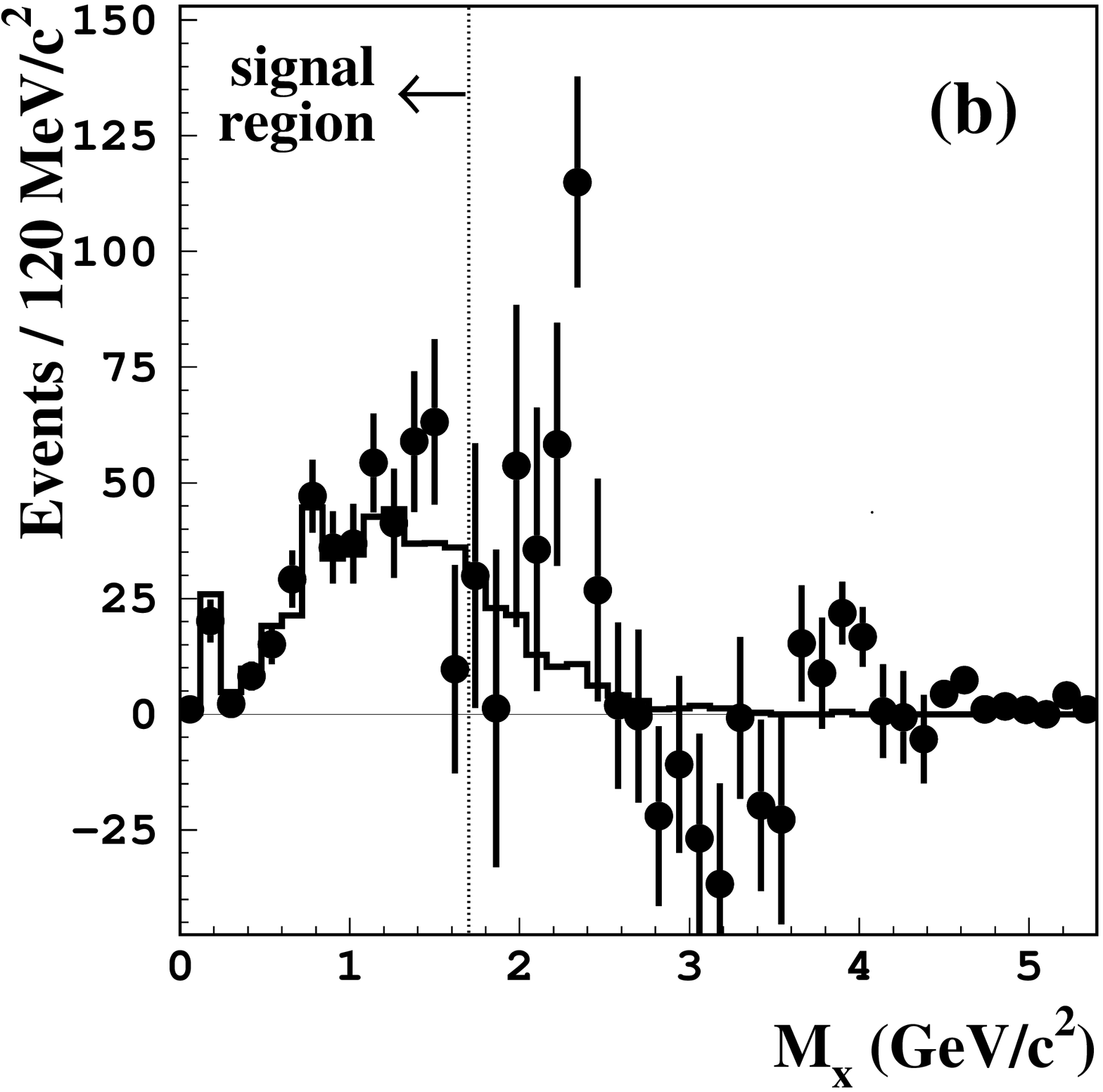}}
\caption{
$M_X$ distribution (no $q^2$ requirement) with fitted
contributions from $X_c \ell \nu$ and $X_u \ell \nu$: (a) before and (b) after
subtracting the $X_c \ell \nu$ contribution (symbols with error bars), shown
with the prediction for $X_u \ell \nu$ (MC, histogram).}
\label{fig:MX-q2}
\end{figure}

Finally, the kinematic variables $M_X$ and $P_+$ are calculated from
the measured momenta of all charged tracks and energy deposits of all
neutral clusters in the electromagnetic calorimeter that are not used
in the \Btag\ reconstruction or for the lepton candidate.
The four-momentum of the leptonic system is calculated as
$q = p_{\ysters} - p_{\Btag} - p_{X}$.
The distributions of events in $M_X$ and $P_+$ are obtained by fitting
the $M_{\rm bc}$ distribution, as described above, in bins of $M_X$
and $P_+$.
Figures~\ref{fig:Mbc-Mxq2}(b), \ref{fig:ppl}(a) and \ref{fig:MX-q2}(a)
show the resulting $M_X$ and $P_+$ distributions. We define three
kinematic signal regions (\sr) for events where the
prompt lepton has \srps: \srpp, {\srMx}, and {\srMx} combined with
\srqd. These three regions are denoted as $P_+$, $M_X$ and $M_X/q^2$,
respectively.
\begin{table}[t]
\begin{center}
\caption{\Nburaw, \epsbu, $F$ and \r\ for the three kinematic signal regions.}
\vspace{0.03in}
\begin{tabular}{lcccc}
\hline\hline
\sr       &~~~\Nburaw~~~&~~~\epsbu~~~&~~~$F$~~~& ~~~\r~~~\\
\hline
$M_X/q^2$ & $268\pm 27$ & $26.5\%$ & $1.03$  & $0.687 \pm 0.014$ \\
$M_X$     & $404\pm 37$ & $28.7\%$ & $1.07$  & $0.700 \pm 0.011$ \\
$P_+$     & $340\pm 32$ & $25.5\%$ & $1.01$  & $0.700 \pm 0.012$ \\
\hline\hline
\end{tabular}
\label{breakdown}
\end{center}
\end{table}
To minimize the systematic effects of uncertainties in lepton
selection and full reconstruction, we normalize the
partial rate for each signal region with the total semileptonic rate:
\begin{equation}
W=\frac{\prate}{\Gamma(X \ell \nu)}=\frac{\Nburaw}{N_{\rm sl}} \times
\frac{F}{\epsbu} \times
\frac{\varepsilon_{\rm frec}^{\rm sl}}{\varepsilon_{\rm frec}^{b \to u}} \times
\frac{\varepsilon_{\ell}^{\rm sl}}{\varepsilon_{\ell}^{b \to u}}.
\end{equation}
To extract the raw number of signal events, \Nburaw, we fit the $M_X$
and $P_+$ distributions with MC-determined shapes for \Bxulnu and
{\Bxclnu} and subtract the {\Bxclnu} contribution. The results for the
$M_X/q^2$ and $P_+$ regions are shown in Figs.~\ref{fig:Mbc-Mxq2}(b)
and \ref{fig:ppl}(a), respectively. MC simulation is used to
estimate the conversion factor $F$ of the observed number of events
\Nburaw\ to the number of signal events produced in the region in
question and observed anywhere, and to estimate the efficiency for
these events, \epsbu.

\Nsl\ is the number of events having at least one lepton with
\srps, determined from a fit to the corresponding $M_{\rm bc}$
distribution (Fig.~\ref{fig:Mbc-Mxq2}(a)), and corrected for the
expected fraction of background events from non-semileptonic decays
(\fake), as estimated by MC simulation.  The factor \rfrec\ accounts
for a possible difference in the \Btag\ reconstruction efficiency in
the presence of a semileptonic or \Bxulnu decay; \rlept\ is the ratio
of both lepton identification efficiencies and fractions of
semileptonic decay leptons with \srps, in the whole kinematic phase
space for semileptonic decays, and within the kinematic signal
region for signal events. The product of efficiency ratios $\r \equiv
\rfrec \times \rlept$ is obtained from MC simulation.
Table I summarizes the results for \Nburaw, \epsbu, $F$ and \r\ for
all three signal regions, where the error in \Nburaw\ is statistical
only. Inserting these values in Eq. 1, we obtain the
three values of $W$. As both numerator and denominator of $W$ have
been obtained from the same tag sample, the $B^0/B^+$ weightings are
the same, and $W$ has no dependence on lifetimes. Multiplying $W$ by
the average measured semileptonic rate $\Gamma(X \ell \nu)={\cal{B}}(B
\to X \ell \nu)/\tau_B$, obtained from \Brsl\ and \tauB~\cite{bib:PDG2004},
gives the average \prate. The results with relative errors are given
in Table II.

\begin{table}[t]
\begin{center}
\caption{Partial rates to the three kinematic signal regions with relative errors (in \%).}
\vspace{0.03in}
\begin{tabular}{lcccccc}
\hline\hline
\sr         & $\prate$    &~~stat~~&~~syst~~&~\btou~&~\btoc~\\ 
\hline
$M_X/q^2$ ~~& $5.24\times 10^{-4}~\mathrm{ps^{-1}}$ & 10.0   & 8.9    & 6.2   & 5.3 \\
$M_X$       & $7.71\times 10^{-4}~\mathrm{ps^{-1}}$ &  9.1   & 7.1    & 6.1   & 2.2 \\
$P_+$       & $6.89\times 10^{-4}~\mathrm{ps^{-1}}$ &  9.4   & 9.3    & 6.4   & 8.7 \\
\hline\hline
\end{tabular}
\label{tab:par}
\end{center}
\end{table}

We divide the experimental error into four categories: statistical,
systematic, \btoc and \btou MC modeling errors, and summarize them in
Table II for the three \prate\ measurements. The two modeling errors
include the uncertainty in signal event extraction, efficiency and
unfolding factor determination due to the choice of specific
theoretical models and values of the parameters used in our MC
predictions.  For signal \Bxulnu MC, the shape function parameters
$\LSF=(0.66 \pm 0.15) \gevcd$ and $\lSF= -(0.40 \pm 0.20) \gevdcd$
were varied within the stated limits, taking into account the negative
correlation between them~\cite{bib:limo-noz}. To take into account the
uncertainty of the prediction in Ref.~\cite{bib:bu-model-fn}, we use a
factor of two larger error for \LSF\ than was determined in
Ref.~\cite{bib:limo-noz}. For \Bxclnu MC, the uncertainty due to our
limited knowledge of branching fractions is studied by varying the
contributions of $D \ell \nu$ and $D^* \ell \nu$ and the relative
fraction of narrow states $D_1$ and $D_2^*$ that contribute to
$D^{**}\ell \nu$ to estimate the modeling error of the $D^{**}$
region. The uncertainty from form factor modeling in $ D \ell \nu$ and
$D^* \ell \nu$ was studied by varying the parameters $\rho^2_D=1.15
\pm 0.16$ and $\rho_A^2=1.56\pm 0.13$ within their
errors~\cite{bib:PDG2004}. The validity of the \Bxclnu simulation was
tested on a \Bxclnu enhanced control sample, where all selection
requirements are applied but with the kaon veto reversed. The
kinematic distributions of this control sample are accurately
described by the simulation.  Other sources of uncertainties, namely
limited MC statistics, extraction of \r, fitting procedure and
imperfect detector simulation are combined in the systematic error.
The uncertainties due to inaccurate simulation of tracking, particle
identification, and cluster finding are estimated by varying for each
source the efficiency within the expected error and taking the maximum
change in \prate\ as the error. For each of these
sources the effects on simulated \btou and \btoc events are
correlated, and the associated shifts are summed linearly.  The net
contributions from the three sources are then summed in quadrature.

The CKM matrix element \Vubs is obtained directly from the partial
rate using \vubformula. $R(\sr)$ is the theoretical prediction of
\prate, the partial rate with a prompt lepton with \srps, divided
by $\Vubs^2$.
The values of $R$ (in $\rm{ps}^{-1}$) are calculated to be \Rmxqd,
\Rmx\ and \Rppl\ for the $M_X/q^2$, $M_X$ and $P_+$ signal regions,
respectively. The $R(\sr)$ values and their errors (SF) are calculated
using the shape function scheme~\cite{bib:generator2} parameters
$\mbSFv$ and $\mupidSFv$ with correlation coefficient $\rho=-0.26$,
obtained from the result of a global fit to moments of both $b \to c
\ell \nu$ and $b \to s \gamma$ distributions \cite{bib:henning}.
While the dependence of $R(\sr)$ on \mupidSF\ is small, we can approximate
the dependence on \mbSF\ as $R/R(m_b^0)-1=k(\sr) \cdot (m_b/m_b^0 - 1)$, where
$m_b^0=4.60~\gevcd$ and $k(\sr)$ is found to be $2.09$, $2.29$ and $3.00$
for the $M_X/q^2$, $M_X$ and $P_+$ signal regions, respectively.
The theoretical error of $R$ (th.) is estimated by varying the
subleading shape functions (four models), the matching scales $\mu_h$,
$\mu_i$, $\bar{\mu}$ and weak annihilation~\cite{bib:generator2}.  The
values of \Vubs with errors are given in Table III. The total error on
\Vubs is 10\%, 9\% and 11\% for $M_X/q^2$, $M_X$ and $P_+$ regions,
respectively. When the shape function parameters and $R$ are better
determined, \Vubs\ can be recalculated from \prate\ shown in Table
II.
\begin{table}

\caption{Values for \Vubs with relative errors (in \%) for the three
kinematic signal regions. Shape function parameters
used in the calculation are $\mbSFv$ and $\mupidSFv$.}
\begin{center}
\begin{tabular}{lccccccc}
\hline\hline
\sr          & $\Vubs \times 10^3$ & ~stat~ & ~syst~ & \btou & \btoc & ~SF~ & ~th.~  \\
\hline
$M_X/q^2$ & $4.70$ & 5.0 & 4.4 & 3.1 & 2.7 & 4.2 & ${}_{\,-5.2}^{\,+4.8}$ \\
$M_X$     & $4.09$ & 4.6 & 3.5 & 3.1 & 1.1 & 4.5 & ${}_{\,-3.8}^{\,+3.5}$ \\
$P_+$     & $4.19$ & 4.7 & 4.6 & 3.2 & 4.4 & 5.8 & ${}_{\,-3.5}^{\,+3.4}$ \\
\hline\hline
\end{tabular}
\label{tab:Vub}
\end{center}
\end{table}

The precision of the \Vubs determination is better than previous
measurements ~\cite{bib:inclusive,bib:Babar_frec,bib:Kakuno}, owing to
the use of larger data sample, better shape function parameter
determination and improved theoretical predictions~\cite{bib:bosch,bib:generator1}.
We find that the usage of the variable $P_+$ is more sensitive to $b
\to c$ modeling and shape function parametrization than the other two
methods and will become competitive in the future when the theoretical
error of $R$ dominates.
No significant experimental nor theoretical improvement was observed
by applying the additional selection \srqd\ to the $M_X$ analysis.
Taking correlations into account, we find that the difference between
\Vubs values for $M_X/q^2$ and $M_X$ regions has a significance of
$2.7 \sigma$.
We conclude that the results are consistent within errors, but we do
not rule out possible effects of duality violation or weak
annihilation contribution.
We chose the $M_X$ signal region result for our \Vubs determination,
since it includes the largest portion of phase space and is least
affected by the uncertainties: \Vubs$=$\Vubv, where the errors
are statistical, systematic with MC modeling, theoretical and from shape function
parameter determination, respectively.
The effectiveness of \Vubs measurements using full reconstruction
tagging is clear (Figs.~\ref{fig:ppl}(b)
and \ref{fig:MX-q2}(b)).

We thank the KEKB group for excellent accelerator operations, the KEK
cryogenics group for efficient operation of the solenoid, and the KEK
computer group and NII for valuable computing and Super-SINET network
support. We acknowledge support from MEXT and JSPS (Japan); ARC and
DEST (Australia); NSFC (contract No.~10175071, China); DST (India);
the BK21 program of MOEHRD and the CHEP SRC program of KOSEF (Korea);
KBN (contract No.~2P03B 01324, Poland); MIST (Russia); MHEST
(Slovenia); SNSF (Switzerland); NSC and MOE (Taiwan); and DOE (USA).

We are grateful to B. Lange, M. Neubert and G. Paz for providing us
with their theoretical computations implemented in an inclusive
generator. We would specially like to thank M. Neubert for valuable
discussions and suggestions.


\begin{thebibliography}{99}

\bibitem{bib:q2cut}
C.~W.~Bauer, Z.~Ligeti and M.~Luke, Phys. Rev. {\bf D 64}, 113004  (2001).

\bibitem{bib:bosch}
S.~W.~Bosch, B.~O.~Lange, M.~Neubert and G.~Paz,  Phys Rev Lett. {\bf 93},221801(2004);
                                              Nucl.\ Phys.\ B {\bf 699}, 335 (2004).
\bibitem{bib:generator1}
M.~Neubert, Eur.\ Phys.\ J.\ C {\bf 40}, 165 (2005);
Phys.\ Lett.\ B {\bf 612}, 13 (2005); hep-ph/0411027;
S.~W.~Bosch, M.~Neubert and G.~Paz, JHEP {\bf 0411}, 073 (2004).

\bibitem{bib:Babar_frec}
B.~Aubert {\it et al.} (BaBar Collaboration), Phys. Rev. Lett. {\bf 92}, 071802 (2004).

\bibitem{bib:Kakuno}
H.~Kakuno {\it et al.} (Belle Collaboration), Phys. Rev. Lett. {\bf 92}, 101801 (2004).

\bibitem{bib:BELLE}
A.~Abashian {\it et al.} (Belle Collaboration), Nucl. Instrum. and Meth. A {\bf 479}, 117 (2002).

\bibitem{bib:KEKB}
S.~Kurokawa and E.~Kikutani, Nucl. Instrum. and Meth. A {\bf 499}, 1 (2003),
and other papers in this Volume.

\bibitem{bib:GEANT}
R.~Brun, F.~Bruyant, M.~Maire, A.~C.~McPherson and P.~Zanarini,
CERN Report No. DD/EE/84-1 (1984).

\bibitem{bib:evtgen}
D.~J.~Lange, Nucl.\ Instrum.\ Meth.\ A {\bf 462}, 152(2001).
 
\bibitem{bib:bu-model-lcsr}
P.~Ball, arXiv:hep-ph/0306251.

\bibitem{bib:bu-model-isgw2}
D.~Scora and N.~Isgur, Phys. Rev. D {\bf 52}, 2783 (1995).

\bibitem{bib:bu-model-fn}
F.~De~Fazio and M.~Neubert, JHEP {\bf 9906}, 017 (1999).

\bibitem{bib:bc-model} QQ event generator, developed by CLEO Collaboration,
see http://www.lns.cornell.edu/public/CLEO/soft/QQ.

\bibitem{bib:HQET-ff}
M.~Neubert,  Phys.\ Rept.\  {\bf 245}, 259 (1994).

\bibitem{bib:generator2} 
B.~O.~Lange, M.~Neubert and G.~Paz, hep-ph/0504071 and private communication with M.~Neubert.

\bibitem{bib:R2}
G.~C.~Fox and S.~Wolfram, Phys. Rev. Lett. {\bf 41}, 1581 (1978).

\bibitem{bib:argus}
H.~Albrecht {\it et al.} (ARGUS Collaboration), Z.~Phys. C {\bf 48}, 543 (1990). 

\bibitem{bib:cball}
J.~E.~Gaiser {\it et al.}, Phys. Rev. D {\bf 34}, 711 (1986).

\bibitem{bib:PDG2004}
S.~Eidelman {\it et al.}, Phys. Lett. B {\bf 592}, 1 (2004).

\bibitem{bib:limo-noz}
A.~Limosani and T.~Nozaki (Heavy Flavor Averaging Group), hep-ex/0407052.

\bibitem{bib:henning}
O.~Buchmueller and H.~Flaecher (Heavy Flavor Averaging Group), hep-ph/0507253.

\bibitem{bib:inclusive}
R.~Barate {\it et al.} (ALEPH Collaboration), Eur. Phys. J. C {\bf 6}, 555 (1999);
M.~Acciarri {\it et al.} (L3 Collaboration), Phys. Lett. B {\bf 436} 174 (1998);
P.~Abreu {\it et al.} (DELPHI Collaboration), Phys. Lett. B {\bf 478} 14 (2000);
G.~Abbiendi {\it et al.} (OPAL Collaboration), Eur. Phys. J. C {\bf 21}, 399 (2001);
A.~Bornheim {\it et al.} (CLEO Collaboration), Phys.\ Rev.\ Lett.\  {\bf 88}, 231803 (2002)

\end{thebibliography}
\end{document}